# Globular bundles and entangled network of proteins (CorA) by a coarse-grained Monte Carlo simulation

Warin Jetsadawisut[1], Sunan Kitjaruwankul[2], Panisak Boonamnaj[1], Pornthep Sompornpisut[1], R.B. Pandey[3*]

[1]*Center of Excellence in Computational Chemistry, Department of Chemistry, Chulalongkorn University, Bangkok 10330, Thailand*
[2]*Faculty of Science at Sriracha, Kasetsart University Sriracha Campus, Chonburi 20230, Thailand*
[3]*Department of Physics and Astronomy, University of Southern Mississippi, Hattiesburg, MS 39406, USA*
[*]*Corresponding author, email: ras.pandey@usm.edu*

Using a coarse-grained model, self-organized assembly of proteins (e.g. CorA and its inner segment iCorA) is studied by examining quantities such as contact profile, radius of gyration, and structure factor as a function of protein concentration at a range of low (native phase) to high (denature phase) temperatures. Visual inspections show distinct structures, i.e. isolated globular bundles to entangled network on multiple length scales in dilute to crowded protein concentrations. In native phase, the radius of gyration of the protein does not vary much with the protein concentration while that of its inner segment increases systematically. In contrast, the radius of gyration of the protein shows enormous growth with the concentration due to entanglement while that of the inner segment remains almost constant in denatured phase. The multi-scale morphology of the collective assembly is quantified by estimating the effective dimension $D$ of protein from scaling of the structure factor: collective assembly from inner segments remains globular ($D \sim 3$) at almost all length scales in its native phase while that from protein chains shows sparsely distributed morphology with $D \leq 2$ in entire temperature range due to entanglement except in crowded environment at low temperature where $D \sim 2.6$. Higher morphological response of chains with only the inner-segments due to selective interactions in its native phase may be more conducive to self-organizing mechanism than that of the remaining segments of the protein chains.

## 1 Introduction

Self-assembly of proteins [1-8] provides mechanical strength and stable dynamical response to the underlying environment such as membrane via its hierarchical morphology. It can provide a reliable and responsive pathways in ion channels [9-23] for selective transport of such specific elements as potassium ions. There are many examples of protein self-assembly with conflicting (adverse and cooperative) effects [3, 4]. For example, the self-assembly of proteins into a viral capsids is critical in preserving the genome from non-conducing external factors such as digesting enzymes of the host cells, undesirable pH, temperature, etc. [3]. On the other hand, the self-assembly of proteins triggered by the conformational changes of the proteins may lead to undesirable aggregation such as amyloid beta-proteins into fibrils. Of a diverse range of proteins with unique and universal response properties, we consider a transmembrane protein CorA [9-22] with a well-defined inner (iCorA) and outer (oCorA) segments [23]. The functional structure of CorA is known to exist as a homo-pentamer [19] to provide coordinated open and close states for the selective transport of $Mg^{2+}$ across the ion channels. Transport of ions depends on the permeation pathways and consequently on the conformation of individual proteins and their interacting network. How CorA proteins assemble collectively into a well-defined morphology is not well understood.

Despite numerous experimental investigations, understanding of the underlying pathways remain highly speculative. Computer simulations provide a viable tool to probe the structural response of proteins and its assembly that may help clarifying such speculative hypothesis. Extensive computer simulations have been recently performed to understand the structure of CorA protein embedded in a model membrane by all-atom MD simulations [19]. Such atomic-scale investigations are insightful in probing the small-scale structural response, however, it is limited to short time scale despite large-scale computer simulations; it is not feasible to examine large-scale structural responses by such approach without severe constraints. For example, the morphology of a set of five proteins with inner segment confined by a nanodisc seems to deform out if they start from a symmetric configuration (see figure 1). Even with a large-scale all-atom MD simulation, it is rather difficult to see a significant dissociation besides fusion of the symmetry. Therefore, a large-scale coarse-grained analysis [23, 24] is needed to augment and clarify the distinctions (see below) if feasible. Perhaps protein-protein interaction may not be enough to direct five CorA proteins into a stable pentamer; other factors such as a nanodisc or membrane matrix may be needed for the stability. Before incorporating additional constitutive components to probe directed assembly, it would be important to understand the self-assembly of the proteins first.

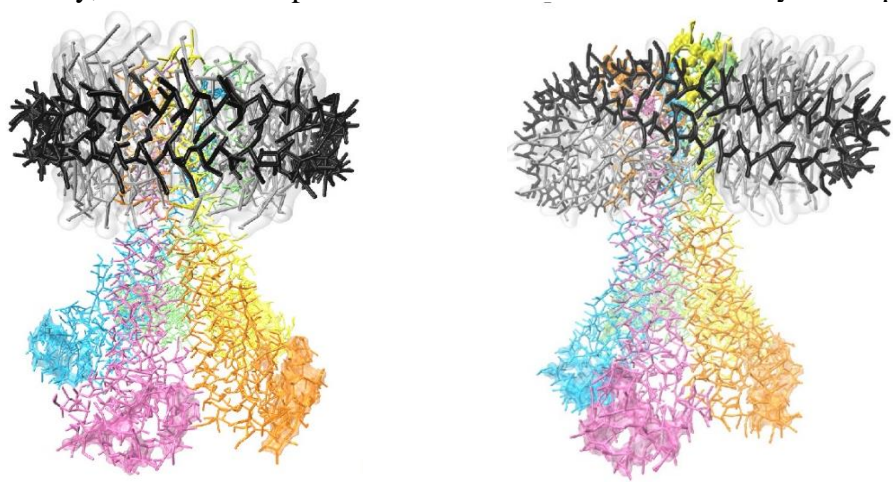

Figure 1: Pentamer of CorA (each protein shown in different colors) with inner core (iCorA) confined by a nanodisc shown in black. The initial configuration is on the left and final configuration on the right after 2 microsecond of coarse-grained MD simulations.

Since the protein channel involves cooperative response of many proteins, it would be desirable to investigate the collective structures in a crowded protein environment. Our goal is to understand the stability of the multi-scale morphologies of interacting proteins in its native and denatured phases. The model is described in the next section, followed by results and discussion with a concluding remark at the end.

**2 Model**

We consider a coarse-grained model [24] of the protein chain on a cubic lattice for simplicity, efficiency, and practical utility: it incorporates specificity of each residue, ample degrees of freedom (unlike minimalist lattice models) for residues to perform their stochastic movements and their covalent bonds to fluctuate. The internal structures of residue (i.e. fine-grained details at atomistic scale) are ignored, however, the specificity of residues is considered via their unique residue-residue interactions (see below) which is critical in capturing the unique characteristics of the protein. In our coarse-grained representation, CorA is a chain of 351 nodes (residues) tethered together on a cubic lattice by flexible peptide bonds. A node represents a residue and occupies a unit cell (of size $(2a)^3$, with lattice constant $a$); the bond length between consecutive residues varies between 2 and $\sqrt{(10)}$ in unit of lattice constant ($a$) [25]. Typically, a protein chain is placed in the simulation box in a random configuration, initially with minimum bond length ($2a$) between the consecutive nodes; this initial configuration is further randomized by allowing each node to perform its stochastic movement (to its 26 neighboring cells with varying distance) with the strictly implemented excluded volume constraint. The bond-fluctuating mechanism has been used extensively in addressing a range of complex problems in polymers [25]; we have extended its utility to model protein chains [23, 24]. Coarse-graining of the simulation box (cubic lattice) and the residues representation by a lattice cell make it one of the efficient computational method for modeling such systems while retaining the potential for fine-graining [26] to further enhance the degrees of freedom.

Each residue interacts with the neighboring residues within a range ($r_c$) of interaction with a generalized Lennard-Jones potential,

$$U_{ij} = \left[ \left| \varepsilon_{ij} \right| \left( \frac{\sigma}{r_{ij}} \right)^{12} + \varepsilon_{ij} \left( \frac{\sigma}{r_{ij}} \right)^{6} \right], \quad r_{ij} < r_c \qquad (1)$$

where $r_{ij}$ is the distance between the residues at site $i$ and $j$; $r_c = \sqrt{8}$ and $\sigma = 1$ in units of lattice constant. We use a knowledge-based residue-residue interaction as input in our phenomenological potential (1) for $\varepsilon_{ij}$. The knowledge-based residue-residue interactions [27, 28] are derived from the distribution of the amino acids in a growing ensemble of frozen protein structures in protein data bank (PDB); the underlying solvent environment is therefore taken into account implicitly. Various assumptions and approximations [27, 28] are further made in deriving these contact potentials which makes it somewhat difficult to calibrate the scales of the physical quantities in absolute units. The relative strength of residue-residue interactions is critical for protein to adopt its specific conformations. Thus, the potential strength, $\varepsilon_{ij}$, is unique for each interaction pair with appropriate positive (repulsive) and negative (attractive) values selected from the knowledge-based contact interactions [27, 28]. The knowledge-based residue-residue interactions have been used extensively in modeling protein structures for decades [29–32]. We use the residue-residue contact matrix by Betancourt-Thirumalai (BT) [27], an improved version of classic Miyazawa-Jernigan (MJ) interaction [28]. It is worth pointing out that there can be alternate methods to incorporate the specificity of residue interactions in a phenomenological potential (1) including the computed interactions involving all-atom MD simulations [33, 34].

Structural evolution of protein chains (CorA) are analyzed in detail as they interact, associate, and dissociate in their native (relatively low temperature) to denatured (high temperature) phases. The sample is prepared by inserting protein chains randomly first in the simulation box; protein chains are moved around with excluded volume constraints to randomize their conformation and distribution further. The protein chain CorA consists of *351* residue, inserting many chains, each of length ($L_c = 351$) in a simulation box of size $L^3$ ($=350^3$) become difficult beyond a certain number ($Nc = 5 - 100$) of chains due to onset of jamming. The computer processing unit (CPU) also increases with the number of chains. Therefore, the number of protein chains is restricted to perform computer simulations with reasonable CPUs (order of days to week) while capturing the cooperative and competing effects of residue-residue, segmental, and overall protein-protein interactions on the local and global association and dissociation of protein chains. Further, relaxation time to reach steady-state (by examining the temporal evolution of various physical quantities) also increases which makes it difficult to assess the thermal response of physical quantities in equilibrium. The protein concentration is defined by the fraction *p* of the lattice sites occupied by the residues of the protein chains, i.e. $p = 8 \times L_c / L^3$. The number of protein chains $Nc = 5 - 100$ each with length $L_c = 351$ (protein chain of CorA) and $L_c = 61$ (inner segment of protein) are considered with simulation box of size $L = 350$, and $150$ respectively. Note that the simulation box becomes crowded at a lower protein concentrations with longer chain lengths.

Each residue performs its stochastic movements with the Metropolis algorithm which involves selecting a residue randomly at a site say *i* of a randomly selected protein chain and attempting to move it to one of its neighboring site say *j*. If the excluded volume constraints and limits on the bond length for the proposed move are satisfied then we calculate the corresponding change in energy $\Delta E_{ij} = E_j - E_i$ between its old ($E_i$) and new ($E_j$) configurations, and move the node with the Boltzmann probability $exp(-\Delta E_{ij}/T)$ where T is the temperature in reduced unit of the Boltzmann constant ($k_B$). Attempts to move each node once defines the unit Monte Carlo step (MCS), i.e. $N_c \times L_c$ attempts to move randomly selected nodes in a simulation box with $N_c$ protein chains each with $L_c$ residues defines the unit MCS. Simulations were performed for sufficiently long time steps (typically 10 million time steps) to make sure that system has reached its steady state equilibrium with a number of independent samples (10-100) to estimate the average values of the physical quantities. A number of local and global physical quantities were determined such as the energy of each residue and protein chain, contact map, their mobility, mean square displacement of the center of mass of the protein, radius of gyration, and its structure factor. These physical quantities are in arbitrary units, i.e., the length is in units of the lattice constant which is different from many all-atom simulations where realistic units for size and time scales are used via calibration of $\sigma$ and $\varepsilon_{ij}$ from experimental data for simplified systems. It is difficult to quantify physical quantities in absolute units due to the phenomenological nature of the interactions (Eq. (1)). The trend in response of the physical quantities to such parameters as the temperature and network density should however be qualitatively comparable with appropriate observations.

**3 Results and discussion**

We study the effects of protein-protein interaction on a number of local and global physical quantities as proteins organize, associate and disassociate as a function of temperature and their concentration (i.e. the number of proteins). Our system has reached the steady-state equilibrium during the course of simulations (i.e. in $10^7$ time steps) for almost entire temperature regime and concentrations except in extreme limits (i.e. high temperature and higher concentrations) where the relaxation time is too large due to entanglement and jamming of extended structures of proteins. Nevertheless, it is illustrative to understand the organizing morphologies as the protein chains assemble, entangle, and disassociate.

3.1 Snapshots, contact map and profiles

Representative snapshots of CorA and that of its inner segments (iCorA) with 100 protein chains each in the simulation box is presented in figure 2 at a low (T=0.020) and a high (T=0.040) temperature. Both, the protein chains and its inner segments appear to disperse almost uniformly. A closer inspection shows that, in native (low temperature) phase, CorA proteins dissociate more on the scale comparable to the size of a protein chain in general while the clusters of inner segments appear to form isolated globular bundles. The residue-residue interaction is more conducive to agglomeration of chain with the inner segments than that of the full protein (CorA). Corresponding snapshots at a higher temperature (T=0.040) show different morphologies where some degree of phase-separation seem to occur with CorA as the proteins continue to dissociate, expand, and entangle. The spreading of extended inner segments at T=0.040 reduces the segregation of protein clusters seen at the low temperature. Attempts are made to quantify our qualitative observations of complex morphologies by analyzing the structure factor (see below).

Contact maps (figures S1-S4) of the assembly of CorA chains and inner segments at a low temperature (T=0.020) seem to suggest a more compact morphology of iCorA in crowded protein environment than that of CorA. At high temperature (T=0.040), frequency of segmental contacts is much lower in their crowded protein matrices than that at the low temperature; segmental contacts in chains with inner segment is still higher than that with CorA. The average number ($N_n$) of residues around each is a measure of the contact profile which is presented in figure 3 and 4 for a range of temperatures (T=0.020–0.040). Obviously, the residue contact density decreases on increasing the temperature with more contacts at low temperatures (native phase) and less in denatured phase. Variation in pattern of the contact profiles with the temperature however, reveals how the protein segments organize during the self-assembly that leads to a global morphology. At the low temperature (T=0.020), most part of the protein chains seem to be surrounded by proteins leading to a compact (globular) morphology. Onset of preferential contacts emerges on raising the temperature (T=0.025). For example, second half of residues in chains with only inner segment (iCorA) of the protein (residues $^{316}$F – $^{351}$L) have higher contact density than the first half (residues $^{291}$M – $^{315}$N) (figure 4). Preferential contact becomes more localized ($^{316}$F – $^{325}$W) on increasing the temperature further (T=0.030). Even though the contact density of many residues is relatively high in self-assembled morphology in CorA, its random distribution along the backbone makes less defined at higher temperatures (figure 3). Further, the fraction of segments with relatively low contact density remains appreciably high. Thus, the protein-protein interaction in inner segments is likely to provide a more stable support for a collective morphology of CorA protein.

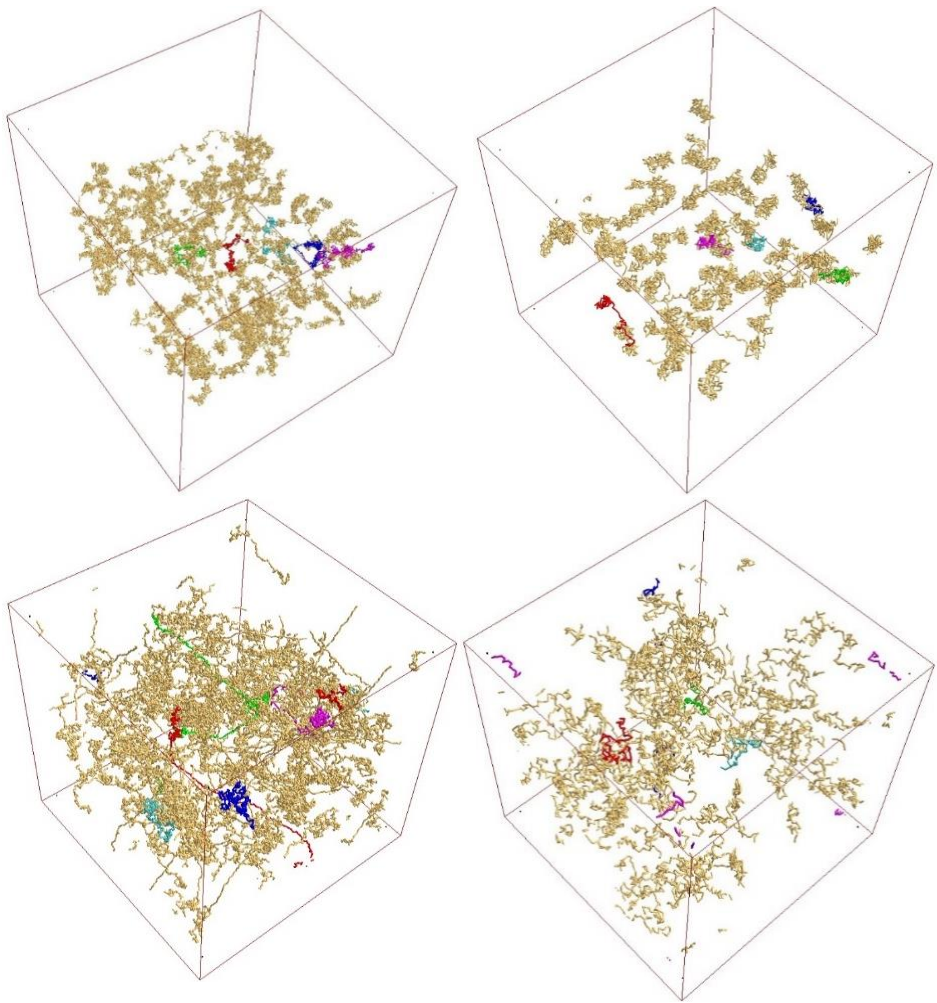

Figure 2: Snapshots of CorA (left column) and iCorA (right column) with 100 protein chains at a low (T=0.020, top) and a high (T=0.040, bottom) temperature at the end of $10^7$ time steps. Five proteins are shown in different colors in each simulation box, remaining (95 chains) are golden khaki.

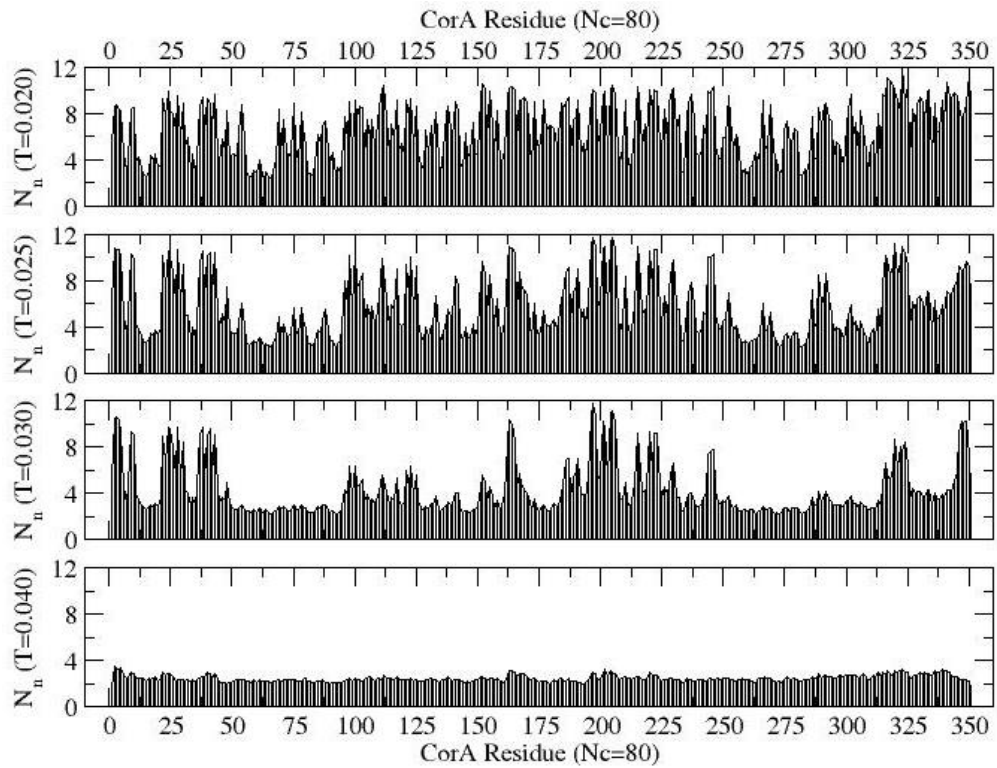

Figure 3: Average number of residues around each of CorA ($^1$M $^2$E ... $^{351}$L) residues in a crowded matrix with 80 chains at low (T = 0.020, 0.025) and high (T = 0.030, 0.040) temperatures.

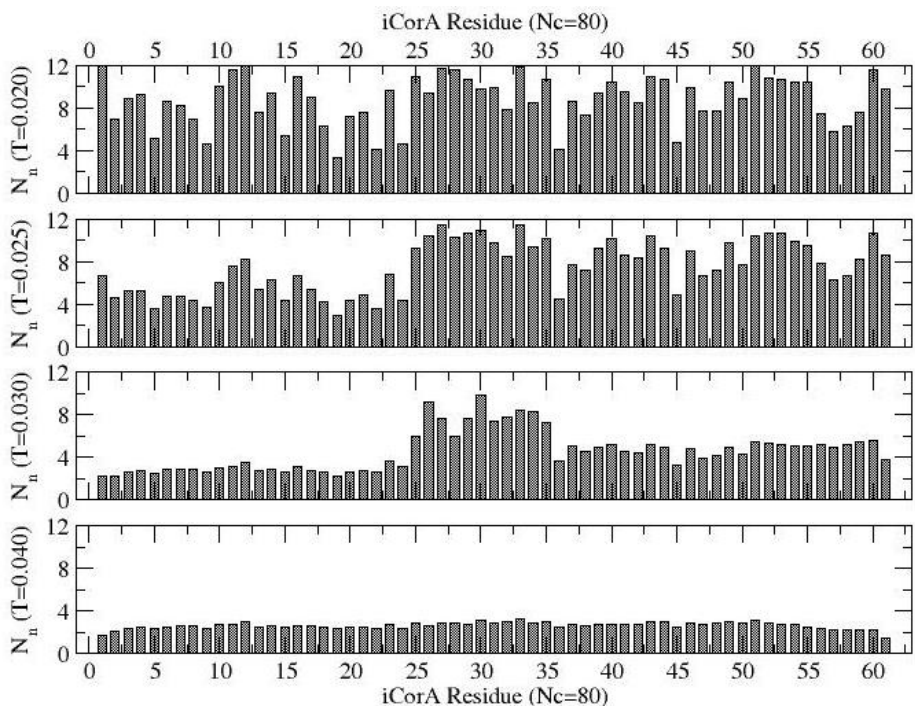

Figure 4: Average number of residues around each of iCorA ($^{291}$M $^{293}$V ... $^{351}$L) residues in a crowded matrix with 80 chains at low (T = 0.020, 0.025) and high (T = 0.030, 0.040) temperatures.

3.2 Radius of gyration

How does the size of protein chains depend on the protein concentration in native and denatured phases? How are residues distributed as the protein chains associate and dissociate? We address it by analyzing the radius of gyration of each protein and structure factor of the self-organized assembly. Variations of the average radius of gyration of the protein CorA with the concentration is presented in figure 5 at representative low and high temperatures. The radius of gyration of the protein does not show much variation with the protein concentration at low temperatures (inset figure 5). At the high temperature (T=0.040) the radius of gyration grows continuously with the concentration reaching a steady-state value in the crowded regime. It should be pointed out that the entanglement is enhanced at higher protein concentrations. We believe that a large increase of the radius of gyration is caused by the competition between entanglement and the thermal expansion of chains.

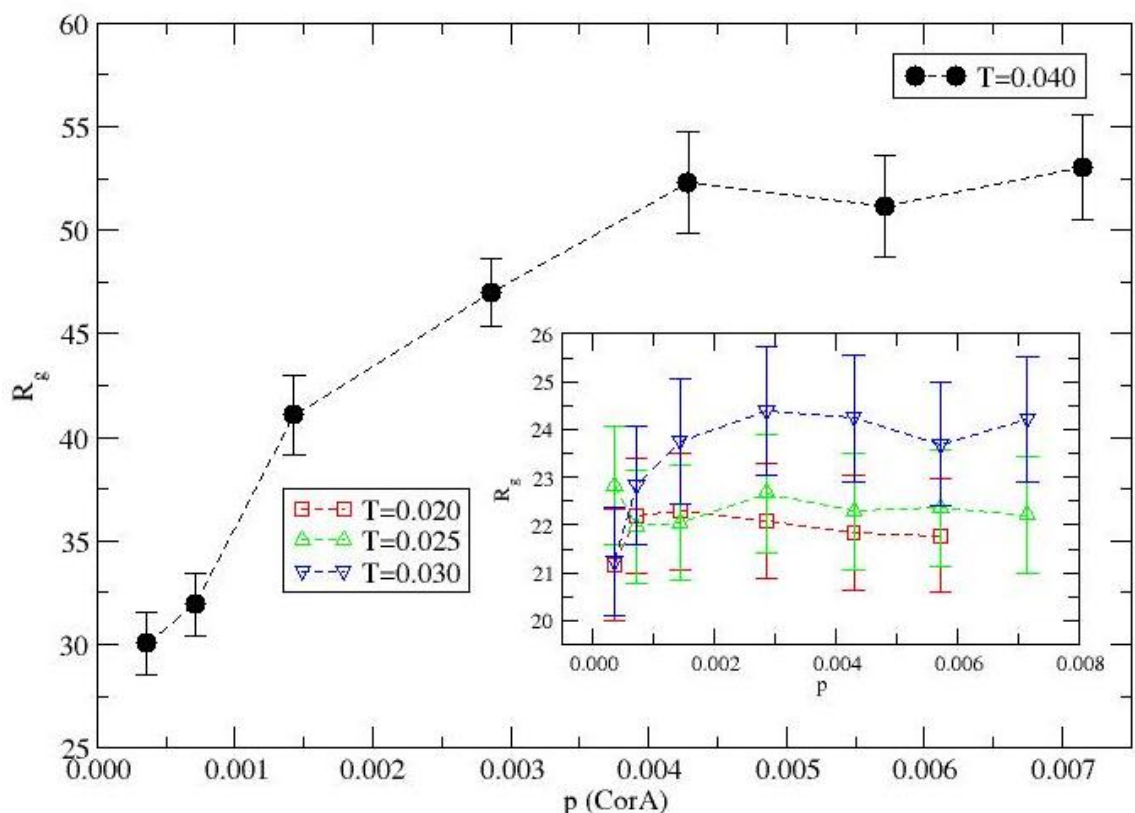

Figure 5: Variation of the average radius of gyration $R_g$ of CorA with the protein concentration p at temperature T = 0.020-0.040.

Figure 6 shows that the radius of gyration of chains with only inner segments increases with the protein concentration p systematically at low temperatures (T=0.015, 0.020) in its native phase and exhibits little variation with p at higher temperatures (T $\geq$ 0.025) in its denatured phase. We therefore believe that the protein-protein interaction among the inner-segments of the protein in its native phase may be more responsive to self-organized collective morphology of proteins.

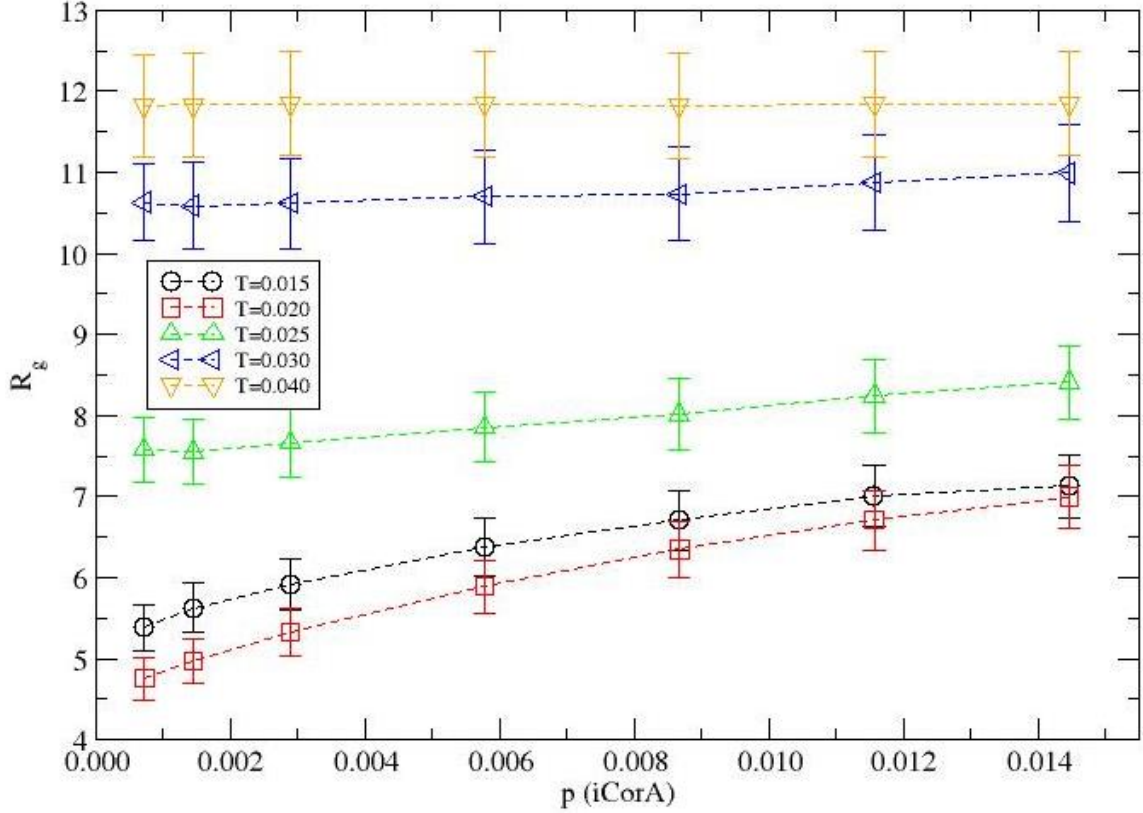

Figure 6: Variation of the average radius of gyration $R_g$ of iCorA with the protein concentration p at temperature T =0.015-0.040.

3.3 Structure factor

Overall distribution of residues over multiple length scales can be assessed by analyzing the structure factor $S(q)$ which is defined as,

$$S(q) = \langle \frac{1}{N} \left| \sum_{j=1}^{N} e^{-i\vec{q}\cdot r_j} \right|^2 \rangle_{|\vec{q}|} \qquad (2)$$

where $r_j$ is the position of each residue in all protein chains and $|q| = 2\pi/\lambda$ is the wave vector of wavelength $\lambda$. Using a power-law scaling of the structure factor with the wave vector, i.e.,

$$S(q) \propto q^{-1/\gamma} \qquad (3)$$

one may study the spread of residues over the length scale $\lambda$ by evaluating the exponent $\gamma$ which describes the mass (residue) distribution. Since we know the overall size of each protein chain via its radius of gyration (figure 5, 6), we can estimate the dependence of the number of residues on multiple length scales ($\lambda$), i.e. from the size of a residue to the linear scale ($L$) of the simulation box. On the spatial length scale comparable to radius of gyration ($\lambda \sim R_g$), we can estimate the scaling exponent $\nu$ from the power-law, $R_g \propto N^\nu$, where $N$ is the number of residues (a measure of the molecular weight of the protein); the effective (fractal) dimension (*D*) of each protein $D = 1/\nu$, $\gamma = \nu$. Similarly we can estimate the mass distribution of the protein self-assembly at $\lambda > R_g$.

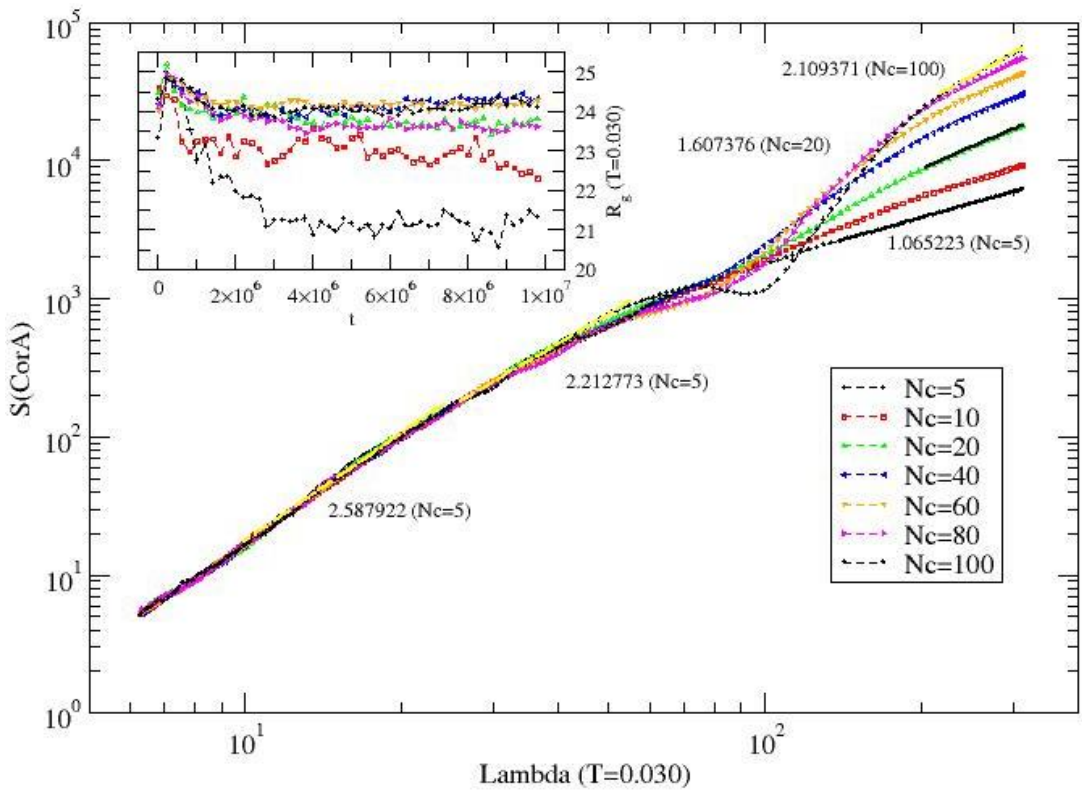

Figure 7: Structure factor *S(q)* versus the wave length (Lambda $\lambda$) at temperature T =0.030 with a wide range of the number Nc (5-100) of CorA proteins in simulation box. The inset shows the variation of the radius of gyration of proteins with the number Nc (5-100) of CorA proteins in simulation box.

Variation of the structure factor (*S(q)*) with the wave length ($\lambda$) of the self-assembly of the protein (CorA) is presented in figure 7 at temperature T = 0.030 for a wide range of protein concentrations, dilute (Nc=5) to crowded (Nc=100) regime. Note that the radius of gyration has reached a steady-state value for all concentrations of the protein at T = 0.030. We see that D ~ *2.6* for each protein ($\lambda \sim R_g$), D ~ 2.2 at $R_g \leq \lambda \leq 50$ (in unit of lattice constant). For all residues distributed over the entire simulation box, $\lambda \sim L$, the effective dimension decreases systematically from D ~ *2* with Nc=100 to D ~ *1* with Nc=5 where each protein chain appears isolated. At high temperature (T=0.040), it is difficult to reach steady state during the course of simulations for the radius of gyration of CorA proteins at higher protein concentrations due to entanglements (figure S5). Variation of the structure factor with the wave vector (figure S5) reveals a rather tenuous morphology ($\lambda$ from size of a residue to about third of the simulation box, *L/3* much larger than $R_g$ of protein chains) of the entangled fiberous protein chains at the high temperature with an effective dimension D ~ *1.7*.

The structure factor of residues' distribution in a dilute-to-crowded protein environment (Nc=5 −100) of chains with inner segments is presesnted in figure 8 at a high temperature T = 0.040. Each protein chain appear to maintain globular conformation (D ≥ 3), at $\lambda \leq R_g$; on larger scale $\lambda > R_g$, the density varies with length scale, i.e. D ~ *2.6* and 3.4 with increasing length scale up to about half the size of the simulation box ($\lambda = L/2$). The overall morphology ($\lambda \sim L$), seem to become more tenuous on reducing the number of protein chains; $D \sim 1.6$ with $N_c = 100$. Note that the radius of gyration of the protein increases on increasing the number $N_c$ of protein chains (inset figure 8).

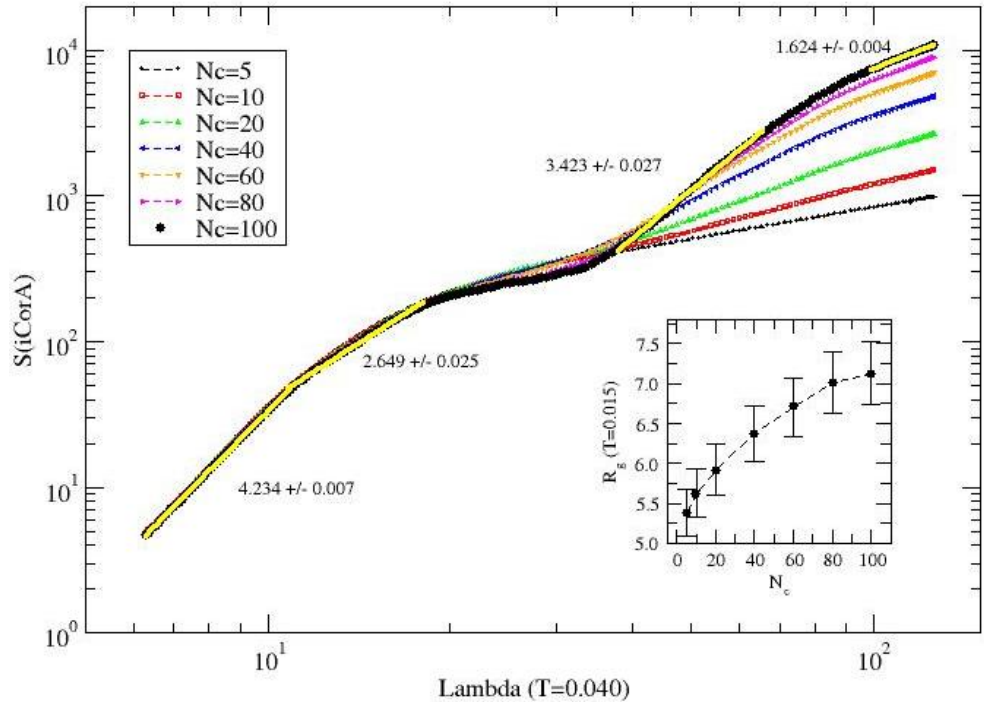

Figure 8: Structure factor $S(q)$ versus the wave length (Lambda $\lambda$) at temperature $T = 0.040$ with a wide range of the number $N_c$ (5-100) of chains with inner segment (iCorA) of the protein in simulation box. The inset figure shows the variation of the radius of gyration of the protein with the number of protein chains.

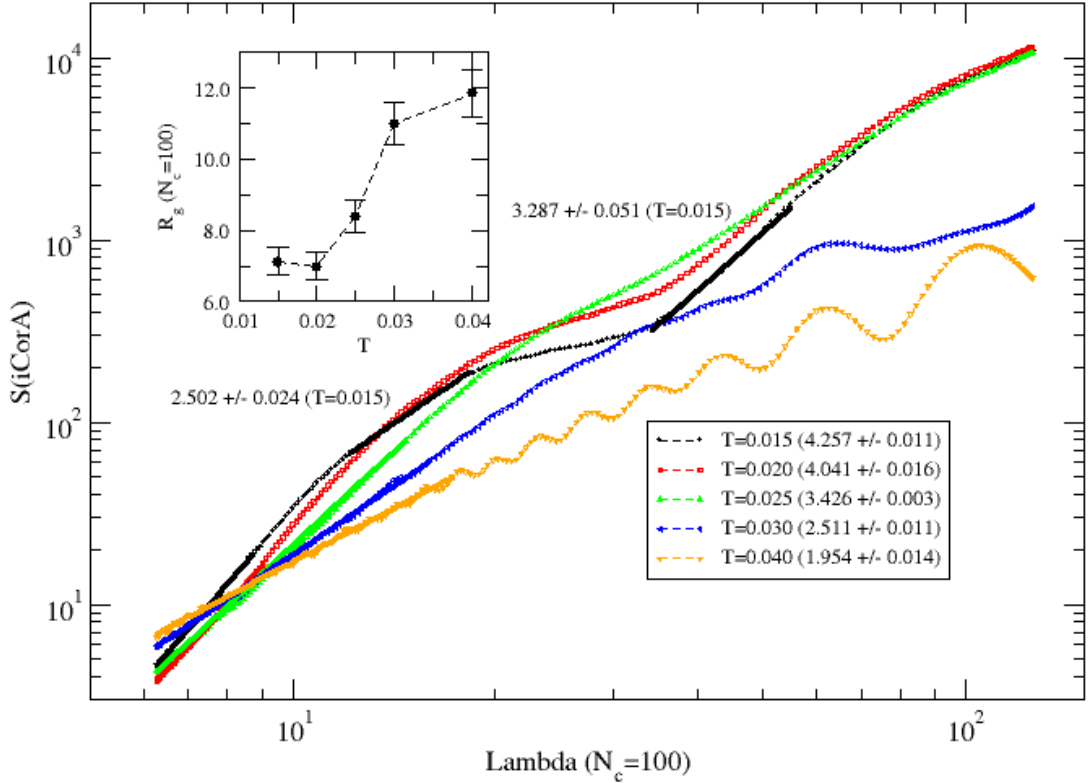

Figure 9: Structure factor $S(q)$ versus the wave length (Lambda $\lambda$) in a crowded protein environment with Nc=100 at a range of low to high temperatures T =0.015 − 0.040 with the estimates of slopes in brackets; slopes of larger scale data at T = 0.015 are also included. Inset is the variation of the radius of gyration of iCorA with the temperature with number Nc = 100 proteins in simulation box.

The variation of the structure factor with the wavelength in the most crowded environment ($N_c=100$) is presented in figure 9 for a range of temperatures. The globular morphology ($D \geq 3$) of individual protein chain seems to persist at lower temperatures (*T=0.015, 0.025*) at length scale comparable to its size ($\lambda \sim R_g$). On a larger-scale ($R_g < \lambda < 3 R_g$), the assembled structure is not as compact ($D \sim 2.5$), but the overall assembly remains compact (solid with $D \sim 3$) for its spread over the simulation box at the low temperature (T=0.015). The multi-scale morphology of self-organizing proteins is heterogenious regardless of its globular conformations in its native phase. The morphology remains relatively dense $D \sim 2.5$ even at a much higher temperature, i.e. *T=0.030*. On large length scale, ($\lambda \sim L$), the self-assembled morphology appears tenuous. The morpholgy adopts to a random coil $D \sim 1.8$ at high temperature $T = 0.040$ where onset of a regular long range structure appear to sets in (with an oscillation in $S(q)$).

## 4 Conclusions

Self-organized structures of interacting proteins (CorA) and its inner segments are investigated by a coarse-grained Monte Carlo simulation as a function of protein concentration at a range of low to high temperatures. Visual inspection show clear distinctions in morphology of the assembly in dilute solution and that in the crowded (dense) matrix at both low and high temperatures. CorA proteins seem to dissociate more on the scale comparable to the size of the proteins while the clusters of chains with inner segments phase-separate in its native phase (low temperatures). It appears that the protein-protein segmental interaction is more conducive to agglomeration of inner segments than that of the outer segments of the protein.

Variation in pattern of the contact profiles with the temperature reveals how the protein segments organize during the self-assembly that leads to a global morphology. We find that a relatively lower fraction of residues in outer segments (residues $^1$M-$^{290}$V) of the protein participate in segmental globularization (with a random distribution) in comparison to that of the inner core segment ($^{291}$M... $^{351}$L) with well-defined globular region. The radius of gyration of the protein does not vary much with the protein concentration at low temperatures. However, it shows enormous growth as the entanglement increases with the concentration at high temperature. In contrast, the radius of gyration of chains with inner segments increases with the protein concentration p systematically at low temperatures (T=0.015, 0.020) in its native phase. We therefore believe that the protein-protein interaction among the inner-segments of the protein in its native phase is conducive to responsiveness of the self-organizing pathways.

The multi-scale morphology (isolated globular bundles to entangled network) is quantified by evaluating an effective dimension *D* from the scaling analysis of the structure factor with the wave vector. In general, the effective dimension of the assembled morphology of protein chains is much lower than that of its inner segments. For example, the most compact morphology of the proteins with *D ~ 2.6* ($\lambda \sim R_g$) even at the highest protein concentrations (most crowded) appears at low temperature while it remains almost extended in a random configuration with D ≤ 2 at almost entire temperature and concentration range considered here. In contrast, the self-assembled morphology of chains with inner segments remains globular *D ~ 3* at almost all length scales in its native phase. It is difficult to identify main cause of a stable pentamer configuration due to complexity of the crowded proteins resulting from inter- and intra-chain residue-residue interactions, competition between the self-organizing mechanism and steric constraints, and entanglement. Based on our analysis presented here, we believe that a stronger protein-protein interactions among the inner segments of the protein is conducing to its collective stable self-assembly.

ACKNOWLEDGMENTS :This research has been supported by the Ratchadaphiseksomphot Endowment Fund, Chulalongkorn University to PS, the Chulalongkorn university dusadi phipat scholarship to WJ. Support from the Chulalongkorn University for a visiting professorship is gratefully acknowledged by RBP along with the warm hospitality by the Department of Chemistry. The authors acknowledge HPC at The University of Southern Mississippi supported by the National Science Foundation under the Major Research Instrumentation (MRI) program via Grant # ACI 1626217.

# Supplementary material

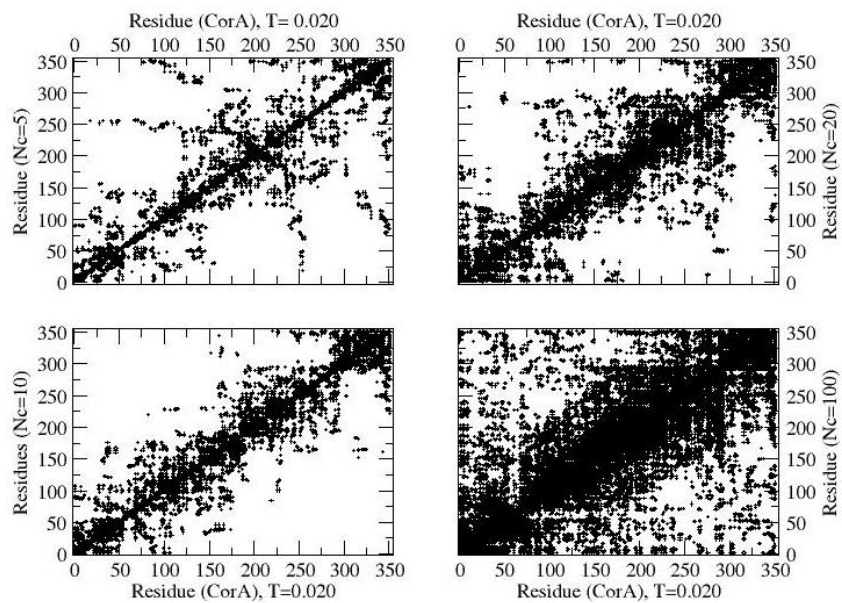

Figure S1: Contact map of residues around each of CorA ($^1$M $^2$E … $^{351}$L) residues in a simulation box with the number of protein chains Nc = 5, 10, 20, and 100 at a low temperature (T = 0.020).

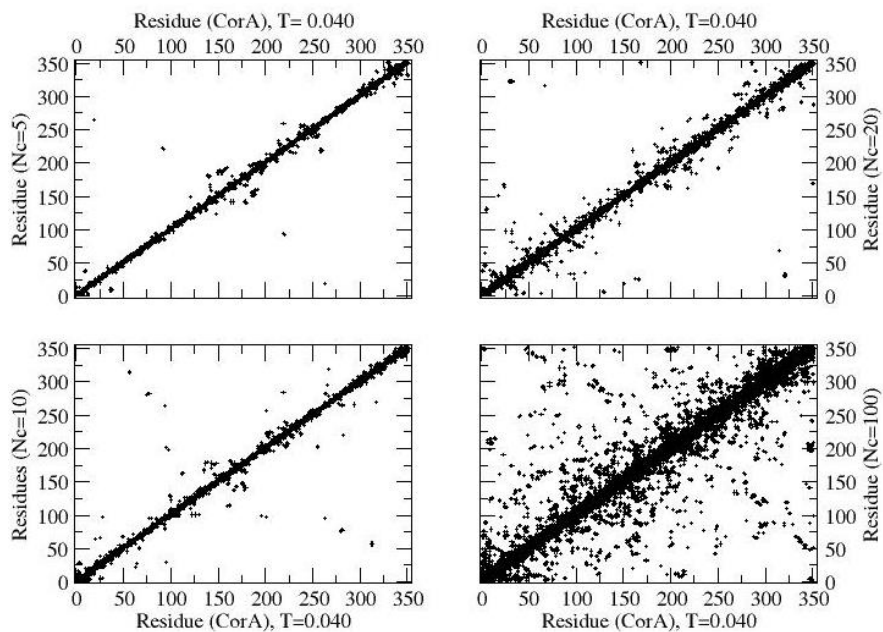

Figure S2: Contact map of residues around each of CorA ($^1$M $^2$E … $^{351}$L) residues in a simulation box with the number of protein chains Nc = 5, 10, 20, and 100 at a high temperature (T = 0.040).

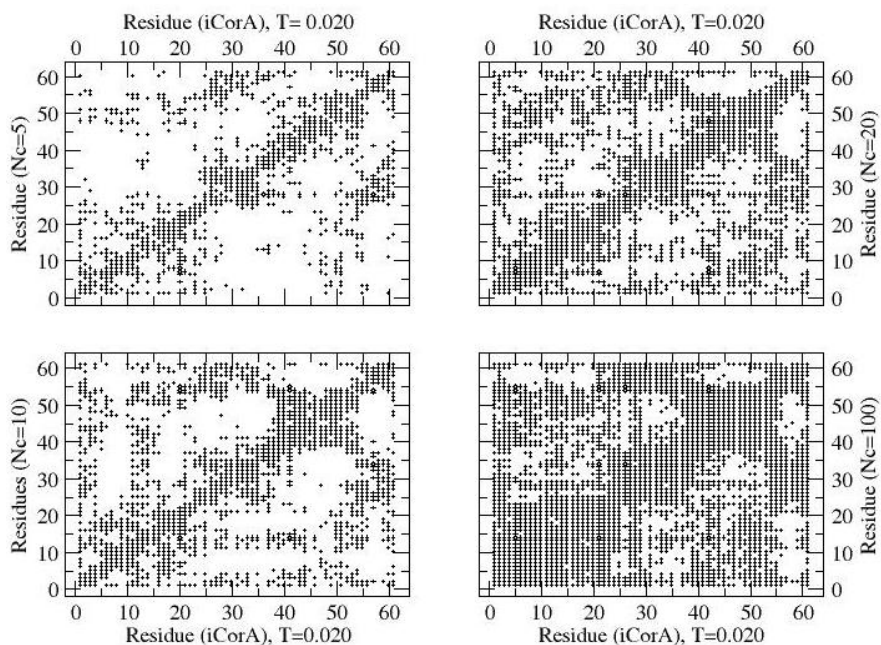

Figure S3: Contact map of residues around each of iCorA residues ($^{1}$M $^{2}$V … $^{61}$L, corresponding sequence in CorA $^{291}$M $^{293}$V … $^{351}$L) in a simulation box with the number of protein chains Nc = 5, 10, 20, and 100 at a low temperature (T = 0.020).

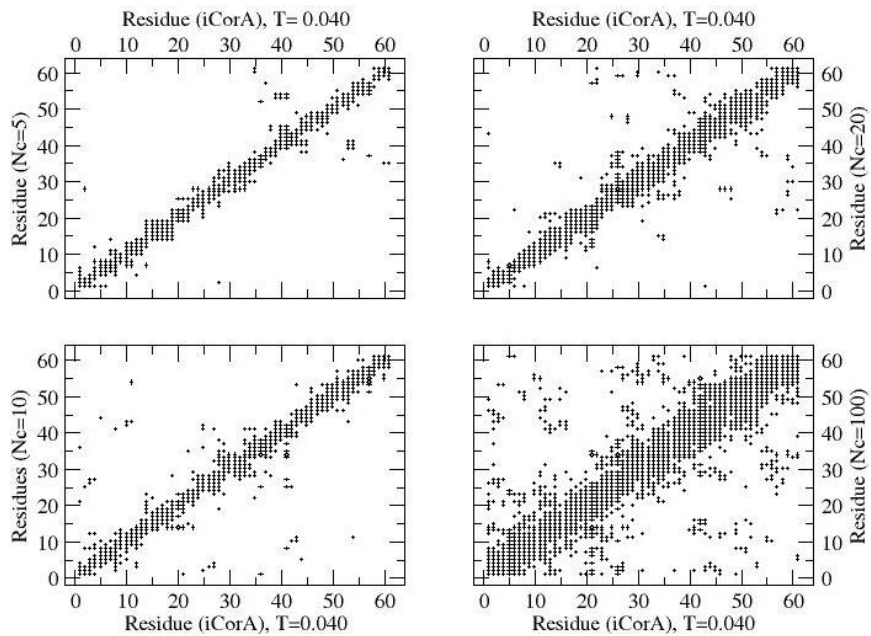

Figure S4: Contact map of residues around each of iCorA residues ($^{1}$M $^{2}$V … $^{61}$L, corresponding sequence in CorA $^{291}$M $^{293}$V … $^{351}$L) in a simulation box with the number of protein chains Nc = 5, 10, 20, and 100 at a high temperature (T = 0.040).

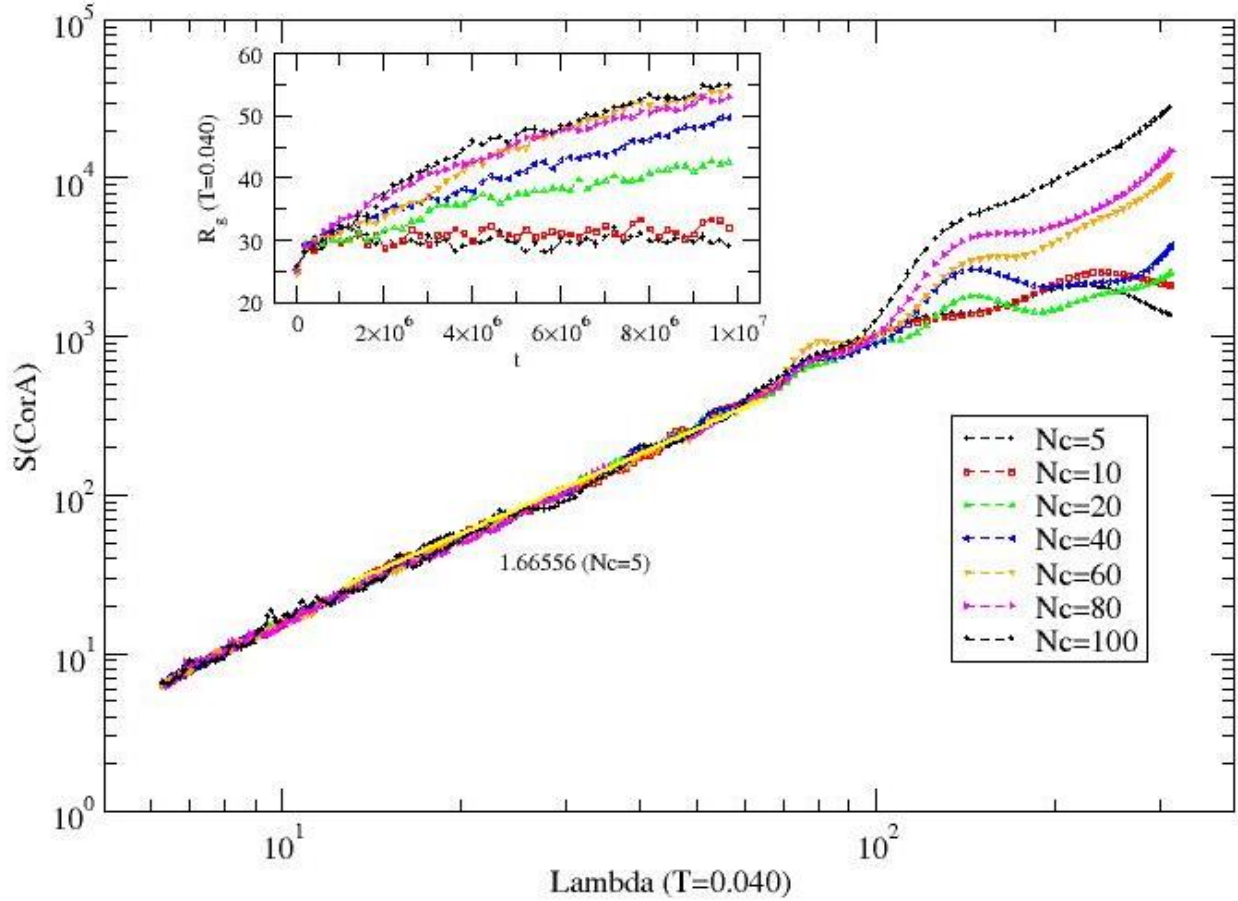

Figure S5: Structure factor *S(q)* versus the wave length (Lambda $\lambda$) at temperature T =0.040 with a wide range of the number Nc (5-100) of CorA proteins in simulation box. The inset shows the variation of the radius of gyration of proteins with the number Nc (5-100) of CorA proteins in simulation box.

18